\documentclass[12pt]{article}
\usepackage{graphicx}
\usepackage{subfigure}
\usepackage{amssymb}
\usepackage{amsfonts}
\usepackage{latexsym}
\usepackage[dvips]{color}

\input epsf.tex

\newcommand{\beq}{\begin{equation}}
\newcommand{\eeq}{\end{equation}}
\newcommand{\be}{\begin{equation}}
\newcommand{\ee}{\end{equation}}
\newcommand{\bea}{\begin{eqnarray}}
\newcommand{\eea}{\end{eqnarray}}

\makeatletter

\@addtoreset{equation}{section}
\makeatother

\begin{document}

\baselineskip=15.5pt
\pagestyle{plain}
\setcounter{page}{1}

\begin{titlepage}
\begin{flushleft}
       \hfill                       FIT HE - 17-01 \\
       \hfill                       
\end{flushleft}

\begin{center}
  {\huge Holographic Cosmology and   
   \vspace*{2mm}
Phase Transitions of SYM theory 
\vspace*{2mm}
}
\end{center}

\begin{center}

\vspace*{5mm}
{\large 
${}^{\dagger}$Kazuo Ghoroku\footnote[1]{\tt gouroku@fit.ac.jp},
${}^{\ast}$Ren\'e Meyer\footnote[3]{\tt Rene.Meyer@physik.uni-wuerzburg.de},
and ${}^{\P}$Fumihiko Toyoda\footnote[4]{\tt ftoyoda@fuk.kindai.ac.jp}
}\\

\vspace*{2mm}
{${}^{\dagger}$Fukuoka Institute of Technology, Wajiro, 
Higashi-ku} \\
{Fukuoka 811-0295, Japan\\}
\vspace*{2mm}
{${}^{\ast}$Institute for Theoretical Physics and Astrophysics, University of W\"urzburg, 97074 W\"urzburg, Germany\\}
\vspace*{2mm}
{ 
${}^{\P}$Faculty of Humanity-Oriented Science and 
Engineering, Kinki University,\\ Iizuka 820-8555, Japan}

\vspace*{3mm}
\end{center}

\begin{center}
{\large Abstract}
\end{center}

We study the time development of 
strongly coupled ${\cal N}=4$ supersymmetric Yang Mills (SYM) theory on cosmological Friedmann-Robertson-Walker (FRW) backgrounds via the AdS/CFT correspondence. 
We implement the cosmological background as a boundary metric fulfilling the Friedmann equation with a four-dimensional cosmological constant and a dark radiation term. 
We analyze the dual bulk solution of the type IIB supergravity and find that the time-dependence of the FRW background strongly influences the dynamical properties of the SYM theory. We in particular find a phase transition between a confined and a deconfined phase. 
We also argue that some cosmological solutions could be related to the inflationary scenario. 

\noindent

\begin{flushleft}

\end{flushleft}
\end{titlepage}

\vspace{1cm}
\section{Introduction}

The holographic approach \cite{ads1,ads2,ads3} 
to supersymmetric Yang Mills (SYM) theory can be extended from the theory on flat Minkowski
space-time to the one on curved boundary background space-times \cite{H,GIN1,GIN2,EGR}. 
Such an extension is useful to understand for example the dynamical properties of the SYM theory on  
cosmologically developing four dimensional (4D) Friedmann-Robertson-Walker (FRW) type space-times. Indeed,   the influence of the cosmological evolution on several dynamical properties of the SYM theory such as e.g. the quark-antiquark potential have been clarified in \cite{H,GIN1,GIN2,EGR}. Since the deconfined phase of QCD behaves in many ways similar to SYM theory at finite temperature, these studies yield interesting information about the behaviour of the quark-gluon plasma in the early universe before big bang nucleosynthesis. 

In the holographic approach, the scale factor $a_0(t)$ is 
undetermined by the bulk equations of motions due to gravity decoupling from the boundary and the boundary metric being nondynamical. For the same reason, the boundary cosmological constant can a priori be any function of time, $\lambda(t)$. In \cite{GIN1,GIN2}, the vacuum state of the SYM theory has been examined by setting $\lambda = (\dot{a}_0/a_0)^2+k/a_0^2$ 
to a constant parameter. This is equivalent to set $a_0(t)$ to be a solution of the 4D Friedmann equation 
with a 4D cosmological constant $\Lambda_4=3\lambda$.
It has been found in particular \cite{GIN1,GIN2} that the ground state of the SYM theory is in the deconfinement and confinement phase for the positive and the negative values of $\lambda$, respectively.  

In \cite{EGR}, this analysis has been extended by including a new parameter $C$, the dark radiation constant,  into the bulk solution. This parameter has been first introduced in \cite{BDEL,Lang} in the context of 
brane world models, and its origin has been discussed in \cite{SMS,SSM}. From the holographic viewpoint, as shown in \cite{EGR}, the bulk solution is related by a large bulk diffeomorphism to the AdS$_5$-Schwarzschild metric with Hawking temperature set by $C$.
The bulk diffeomorphism in particular acts on the boundary metric as a conformal transformation relating the FRW boundary metric to flat Minkowski space-time. 
Its field theory dual therefore represents a finite temperature SYM state in a FRW universe. 
In general, 
the dynamical properties of the dual theory 
are now controlled by the two parameters, $\lambda$ and $C$. 
{In \cite{GN13}, the analysis was extended to time-dependent $\lambda(t)$ by assuming  
the existence of the matter other than the 4D cosmological constant $\Lambda_4$ in the boundary space-time, 
however the time-dependent properties of the SYM state has not
been discussed there.} 
It was found \cite{GN13} that the system is in the deconfinement phase when $C$ is large enough
even if the boundary cosmological constant $\lambda$ is negative. Hence the two parameters $C$ and $\lambda$ can  provide opposite dynamical effects in the theory,  
as shown in \cite{GN13,GITT15}. 

In this work we extend the analysis to the case where 
we can see how the state of the SYM system varies with the cosmological development of our universe. 
Our purpose is to propose a self-consistent
procedure how to do it and to find the time dependence of the state of the SYM theory.
The procedure is as follows: First, we obtain the energy momentum tensor of the SYM theory, $\langle T_{\mu\nu}^{YM}\rangle$, 
in the FRW boundary metric from holographic renormalization \cite{FG,KSS,BFS}. 
Then by using this tensor, the 4D Einstein equations (i.e. the Friedmann equation for the at that point arbitrary scale factor $a_0(t)$)  with
$\Lambda_4$ and SYM as a matter are solved to obtain the boundary scale factor 
$a_0(t)$. After that, the time development of the SYM state is then analyzed by substituting this $a_0(t)$ back into the bulk solution. We notice that the 4D Einstein equations used here were obtained in \cite{AST} 
from a different holographic regularization procedure by adding a 4D boundary 
gravitational action. This formulation may be related also to the designer gravity approach of \cite[Kiri] to couple two AdS spacetimes together such that at the boundary a gravitational theory is induced.


The solutions of $a_0(t)$ obtained in this way provide a "sudden" singularity \cite{Awad} 
in the 4D boundary space-time at the minimum value 
of the scale factor $a_0^{min}$. The boundary cosmological constant $\lambda$ is found as a two valued function of $a_0$ for $a_0>a_0^{min}$. {The two branches merge at the minimal value.} 
We then study the solutions 
for the two cases $\Lambda_4 <0$ and $\Lambda_4 >0$ separately.  

The region of negative $\lambda$ \footnote{We should not confusing $\lambda$ with $\Lambda_4$. It seems they should be the same. We could see their exact difference and relation in the sction three.} 
is realized in the case of $\Lambda_4 <0$, where we observe the  expected
phase transition between Wilson loop confinement and deconfinement. The equation of state $w=p/\rho$, where $p$ and $\rho$ 
denote the pressure and energy densities of the SYM fields, is found to be a useful parameter for this the transition. The parameter $w$, which varies with time, characterizes 
the state of the SYM theory. We found that the critical point of the confinement-deconfinement
transition is at exactly $w=-1/3$. We speculate on why this point is critical in terms of the
Virial theorem. For $\Lambda_4 >0$, on the other hand, inflation behavior will be found 
through this analysis and issues related to their cosmology are discussed. 

As for the second branch of the solutions, which cover a larger positive region of $\lambda$, 
a simple behavior of the solution $a_0(t)$ 
is obtained, and the relation to the cosmology is discussed.  
In both cases, the value of $a_0(t)(>a_0^{min})$ is bounded from below when the dark radiation $C$ is present. 
This implies that we need to take into account quantum gravitational effects to approach the dynamics 
in the region of $a_0<a_0^{min}$ since the curvature diverges at $a_0^{min}$. 
In this region, an approach from quantum cosmology in the mini-superspace would be available to resolve the 
dynamical properties of the system.  

\vspace{.3cm}
The outline of this paper is as follows: 
In the next section, a bulk $M_5\times S^5$ space-time is given as a solution of
10D type IIB supergravity, and previous results for the dual of SYM theory in the FRW space-time
are reviewed. 
In the Sec. 3, the procedure to obtain the cosmological time-development of the
SYM theory by self-consistently calculating $a_0(t)$ is laid out in detail, and the solutions which cross the (de)confinement transition point are studied. Other kinds of solutions and their relation to the inflationary  scenario are discussed in Sec. 4. The solutions, which cover large $\lambda$ region, are given in the Sec. 5, and the relation
to the cosmology is discussed.
Summary and discussions are given in the final, sixth, section.

\section{ Holography of SYM theory in a FRW metric
}


{We start with 
ten-dimensional type IIB supergravity} retaining the dilaton
$\Phi$, axion $\chi$, as well as the selfdual five form field strength $F_{(5)}$,
\beq\label{2Baction}
 S={1\over 2\kappa^2}\int d^{10}x\sqrt{-g}\left(R_{10}-
{1\over 2}(\partial \Phi)^2+{1\over 2}e^{2\Phi}(\partial \chi)^2
-{1\over 4\cdot 5!}F_{(5)}^2
\right). \label{10d-action}
\eeq
All other fields are set to zero, and $\chi$ is Wick rotated to the Euclidean domain \cite{GGP}.
We are going to use the Freund-Rubin
ansatz for $F_{(5)}$, 
$F_{\mu_1\cdots\mu_5}=-\sqrt{\Lambda_5}/2~\epsilon_{\mu_1\cdots\mu_5}$
\cite{KS2,LT} to reduce to the five noncompact dimensions.
 
\vspace{.2cm}
The equations of motion of the above theory are then solved by an Ansatz for  
the 10D metric as $M_5\times S^5$,
$$ds^2_{10}=g_{MN}dx^Mdx^N+g_{ij}dx^idx^j=g_{MN}dx^Mdx^N+R^2d\Omega_{5}^2\, .$$
The five sphere radius $R$ is defined via $1/R=\sqrt{\Lambda_5}/2$.

The holographic dual to the large $N$ gauge theory embedded in 4D space-time with dark energy 
and "dark radiation" is solved by the gravity on the following form of the metric,
\beq\label{10dmetric-2-1}
ds^2_{10}={r^2 \over R^2}\left(-\bar{n}^2dt^2+\bar{A}^2a_0^2(t)\gamma_{ij}(x)dx^idx^j\right)+
\frac{R^2}{r^2} dr^2 +R^2d\Omega_5^2 \ ,
\eeq
where
\beq\label{AdS4-30} 
    \gamma_{ij}(x)=\delta_{ij} \gamma^2(x)\, , \quad \gamma (x)
  =1/\left( 1+k{\bar{r}^2\over 4\bar{r_0}^2} \right)\, , \quad 
    \bar{r}^2=\sum_{i=1}^3 (x^i)^2\, ,
\eeq
and $k=\pm 1,$ or $0$. The arbitrary scale parameter  $\bar{r_0}$ of three space
is set hereafter as $\bar{r_0}=1$.
The solution is obtained from 10D supergravity
of type IIB theory \cite{EGR,EGR2,GN13,GNI13}. 

\bea
 \bar{A}&=&\left(\left(1-{\lambda\over 4\mu^2}\left({R\over r}\right)^2\right)^2+\tilde{c}_0 \left({R\over r}\right)^{4}\right)^{1/2}\, , \label{sol-10} \\
\bar{n}&=&{\left(1-{\lambda\over 4\mu^2}\left({R\over r}\right)^2\right)
         \left(1-{\lambda+{a_0\over \dot{a}_0}\dot{\lambda} \over 4\mu^2}\left({R\over r}\right)^2\right)-\tilde{c}_0 \left({R\over r}\right)^{4}\over
       \sqrt{\left(1-{\lambda\over 4\mu^2}\left({R\over r}\right)^2\right)^2+\tilde{c}_0 \left({R\over r}\right)^{4}}}\, , \label{sol-11}
\eea
where 
\beq\label{Fried-1}
  \left({\dot{a}_0\over a_0}\right)^2+{k\over a_0^2}=\lambda\, . \label{bc-3-1}
\eeq
and
\beq
\tilde{c}_0=C/(4\mu^2a_0^4)\, . \label{sol-12}
\eeq

\begin{itemize}
\item  As for the parameter $C$, at first, this term has been called
as the "dark radiation" 
in the context of the brane world model
\cite{BDEL,Lang}. 
And it has been interpreted as
the projection of the 5D Weyl term \cite{SMS,SSM}. 

\item On the other hand, 
from the holographic
viewpoint, it has been cleared that $C$ corresponds to the radiation energy of the SYM fields \cite{EGR}. 
In the following, we adopt this interpretation. 

\item As for $\lambda $, it is a function of $a_0(t)$ as defined by (\ref{Fried-1}).
Since $a_0(t)$ can't be determined by the equations of motion of 10D bulk gravity, 
then $\lambda $ can be set as an arbitrary function of $t$. Then
for any $\lambda(t)$ and $C$, the above
solution satisfies the 10D Einstein equation of motion given by (\ref{10d-action}).

\end{itemize}

In general, 
we can see how the physical quantities of the SYM theory varies
with time when the t-dependence of $a_0(t)$ has been given. Our purpose is to find it.
Before doing it, we briefly review the results
of the previous analysis where the time dependence of $a_0(t)$ is neglected by assuming
that it varies very slowly. 

\subsection{Phase Diagram for $\lambda =$ const.}

The dynamical properties of the theory
can be studied 
under the assumption that $a_0(t)$ varies very slowly compared to the
time scale of the dynamics of the SYM theory. \cite{EGR}
\footnote{We should notice the following fact that the solution,
$a_0=1/\sqrt{|\lambda|}=$constant, is actually found
for negative constant $\lambda$ and $k=-1$. This is considered as an extremal
case of the slowly varying $a_0(t)$.} 
Since there is no equation to determine $a_0(t)$ in solving the 5D Einstein equations,
then there is no constraint on the time dependence of $a_0(t)$. Then the parameters
appearing in the bulk configuration ($\lambda$ and $b_0$) can be taken as arbitrary values.

In the case of constant $a_0(t)$, the factors $\bar{A}$ and $\bar{n}$ of the above solution are written as follows,
\bea
 \bar{A}&=&\left(\left(1+\left({r_0\over r}\right)^2\right)^2+\left({b_0\over r}\right)^{4}\right)^{1/2}\, , \label{sol-10-1} \\
\bar{n}&=&{\left(1+\left({r_0\over r}\right)^2\right)^2
-\left({b_0\over r}\right)^{4}\over \bar{A}
       }\, , \label{sol-11-1}
\eea
\beq
 r_0=\sqrt{|\lambda|}R^2/2\, , \quad b_0=R\tilde{c}_0^{1/4}\,  . \label{sol-12-1}
\eeq
In this case, the dynamical properties of the dual
4D SYM theory on the boundary are controlled by the $\lambda$ and $C$, or $r_0$ and $b_0$.

\vspace{.3cm}

\begin{figure}[htbp]
\vspace{.3cm}
\begin{center}
\includegraphics[width=8cm]{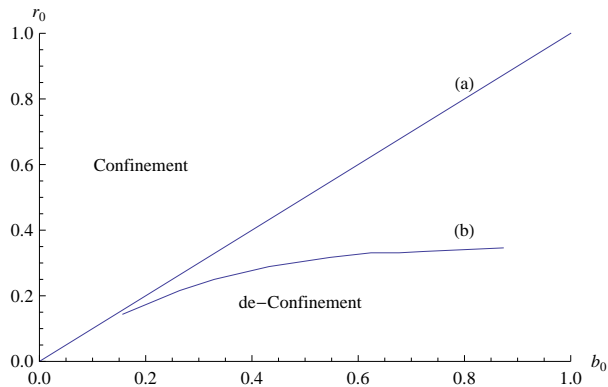}
\caption{ The line (a) shows the critical line $r_0=b_0$ separating the quark-confinement phase and the deconfinement phase 
for the case of constant $a_0(t)$. The curve (b) represents for the running $a_0(t)$ case obtained here. It obeys
the formula $r_0=\left(1+{\dot{\lambda} a_0\over\lambda \dot{a}_0}  \right)^{-1/4} b_0$ which is explained below. }
\label{Phase-diagram}
\end{center}
\end{figure}

\begin{itemize}

\item {\bf $C=0$ and finite $\lambda$:}\\ In this case, 
it has been found
that the SYM theory is in the confinement (deconfinement) phase for negative (positive) $\lambda$. \cite{GIN1,GIN2}.

\item {\bf finite $C$ and $\lambda=0$:}\\
For this case, the 5D metric 
is reduced to the AdS$_5$-Schwarzschild form, in which $C(>0)$ represents the black hole mass.
Then, from the holographic viewpoint, 
$C$ corresponds to
the thermal radiation of the SYM fields at a finite temperature \cite{EGR}. So
the system is in the deconfinement phase. 
 
\item From the above two facts,
we could suppose that the two parameters, $C$ and $\lambda(<0)$,
compete with each other to realize the opposite phase of the theory, namely 
the deconfinement and confinement respectively. In fact,  
we could find the phase transition at a point where these two opposite effects are balanced
\cite{EGR,EGR2,GN13,GNI13,GNI14}. 
The critical line is given by $b_0 = r_0$, where
the density of dark radiation $C$ and the 
magnitude of $|\lambda|$ are replaced by $b_0$ and $r_0$ as
given by the formula (\ref{sol-12-1}). 
The phase diagram of the SYM theory in the FRW space-time is given as in the 
Fig. \ref{Phase-diagram}. 

\item In the deconfinement phase or in the region $b_0\geq r_0$, the temperature $T$ 
is given as the Hawking temperature of the 5D metric. Then $T$ decreases with increasing $r_0$
up to the transition temperature ($T_c$) to the confinement phase. It is given
as $T_c=0$, which corresponds to the critical line $r_0=b_0$ in the Fig. \ref{Phase-diagram}. 

\item For positive $\lambda$ case, the theory is in the deconfinied chiral symmetric
phase. 
The description of the phase in $b_0-r_0$ plane is abbreviated. 

\end{itemize}

\subsection{For the case of time dependent $\lambda$}

The case of the time independent $a_0(t)$ and $\lambda$ is a simple approximation. As shown below, here we
solve the 4D Einstein equation to obtain a time dependence of $a_0(t)$ and $\lambda$ by using the 4D energy
momentum tensor of the SYM theory which is obtained by the holographic method. Then the phase transition line
given by (a) in the Fig.  \ref{Phase-diagram} is modified to (b), 
which obeys the following formula,
\beq\label{critical-L1}
  r_0=\left(1+{\dot{\lambda} a_0\over\lambda \dot{a}_0}  \right)^{-1/4} b_0\, ,
\eeq
The derivation and the details of this result are given below.


\section{Cosmological time development of SYM system}

\vspace{.3cm}
\subsection {Holographic Cosmology}

As mentioned above, the scale factor $a_0(t)$ can not be determined by the bulk equations of motion. 
On the other hand, it is obtained as a solution of 4D cosmological equation, where the gravity 
couples to various matter. 
Here we give it by solving the 4D
Einstein equation which couples to the SYM theory.
In the equations,
we use the energy momentum tensor of SYM, $\langle T_{\mu\nu}^{SYM}\rangle$, 
which is holographically obtained. 
Using the $a_0(t)$ obtained in this way, we can see the time-development of
the state of the SYM theory.

\vspace{.3cm}
Here we consider a simple 4D model, in which
the matter part is dominated by the SYM fields. The action 
to be solved is given as
\beq\label{4d-action}
 S=\int d^{4}x\sqrt{-g}\left\{{1\over 2\kappa_4^2}\left(R_{4}-
      {2}\Lambda_4\right)\right\}+S_{SYM}^{eff}\,  ,\label{4d-action}
\eeq
where $\kappa_4^2=8\pi G_4$ and $\Lambda_4$ denote the 4D gravitational constant and cosmological constant
respectively.
Furthermore, the matter part $S_{SYM}^{eff}$ is written as an effective action which is obtained by integrating
out over all quantum fluctuations of the
SYM fields under the FRW metric, 
\beq\label{RW}
  ds_{(4)}^2=-dt^2+a_0(t)^2\gamma_{ij}dx^idx^j\, .
\eeq

\vspace{.3cm}
The factor $a_0(t)$ is obtained by solving the following Einstein equation,
\beq\label{E2}
  G_{\mu\nu}=R_{\mu\nu}-{1\over 2}Rg_{\mu\nu}+\Lambda_4g_{\mu\nu}
         =\kappa_4^2 \langle T_{\mu\nu}^{YM}\rangle \, ,
\eeq
where 
\beq
      \langle T_{\mu\nu}^{YM} \rangle = {2\over \sqrt{-g}} {\delta S_{SYM}^{eff}\over\delta g^{\mu\nu}}\, ,
\eeq 
which represents the energy momentum tensor of SYM fields. 
It includes all quantum corrections of the interacting SYM fields in the background (\ref{RW}).

As we mentioned in the introduction,
we should notice here that the Eq. (\ref{E2}) has been obtained from the holographic regularization procedure
with a different boundary condition from the Dirichlet condition by adding an appropriate boundary 4D
gravitational action. \cite{AST} 

\vspace{.3cm}
The strategy of our holographic approach is given as follows.

\vspace{.2cm}
\noindent (i) First, $\langle T_{\mu\nu}^{YM}\rangle$ is given by the holographic method.

\vspace{.2cm}
\noindent (ii) Then $a_0(t)$ is obtained by solving the equation (\ref{E2}) for the metric (\ref{RW}).

\vspace{.2cm}
\noindent (iii) Then we examine the time development of the state of the SYM theory  
based on the holographic principle by using (\ref{10dmetric-2-1})-(\ref{sol-12}).

\subsection{Equations to be solved}

The independent equations of (\ref{E2}) are obtained as follows 
\beq\label{bc-RW2}
 \lambda\equiv  \left({\dot{a}_0\over a_0}\right)^2+{k\over a_0^2} = {\Lambda_4\over 3}+{\kappa_4^2\over 3} 
\langle T_{00}^{YM}\rangle\, ,
\eeq
\beq\label{bc-RW3}
 2{\ddot{a}_0\over a_0} +\left({\dot{a}_0\over a_0}\right)^2+{k\over a_0^2} =  {\Lambda_4}-{\kappa_4^2\langle T_{ii}^{YM}\rangle} \, .
\eeq
The first Eq. (\ref{bc-RW2}) represents the $tt$ component of (\ref{E2}), it 
is called as Friedmann equation.

Using the above two equations, we obtain the following continuity equation, 
\beq
 \dot{\rho}+3H(\rho +p)=0\, , \quad  H=\dot{a}_0/a_0\, ,
 \label{continuity-1}
\eeq
where $T_{\mu\nu}^{YM}$ is supposed to be written by $\rho$ and $p$ as
\beq
  \langle T_{\mu\nu}^{YM} \rangle ={\rm diag}(\rho, pg_{ij}^0)\, .
\eeq

The contents of the above $T_{\mu\nu}^{YM}$ are given as follows \cite{GN13,KSS,BFS,FG}
\beq\label{rho}
 \rho=3 \alpha \left( {\tilde{c}_0\over R^4} +{\lambda^2\over 16}\right) \, ,
\eeq
\beq
 p=\alpha\left\{ {\tilde{c}_0\over R^4}-3 {\lambda^2\over 16}\left(
      1+{2\dot{\lambda}\over 3\lambda}{a_0\over \dot{a}_0}\right) \right\}\, ,
 \quad \alpha={4R^3\over 16\pi G_N^{(5)}}\, , \label{density}
\eeq
It is easy to see that the equation (\ref{continuity-1}) is satisfied by the above
$\rho$ and $p$. 

\vspace{.3cm}
We notice that 
to solve the Eqs. (\ref{bc-RW2}) and (\ref{bc-RW3}) is equivalent to solve Eqs. (\ref{bc-RW2}) and Eqs. (\ref{continuity-1}).
On the other hand, the Eq. (\ref{continuity-1}) is satisfied for the above $\langle T_{\mu\nu}^{YM}\rangle$ 
as mentioned above. 
So, our task to do here is to solve the Friedman equation (\ref{bc-RW2}) to obtain $a_0(t)$, and 
the equation to be solved 
is written as
\beq\label{Freed-2}
  \lambda={\Lambda_4\over 3}+{\kappa^2_4\over 3}\langle T_{00}^{SYM}\rangle\
   ={\Lambda_4\over 3}+\tilde{\alpha}^2\left(\tilde{b_0}^4+\lambda^2 \right)\, ,
\eeq
where
\beq
  \tilde{\alpha}^2={\kappa^2_4\over 16}\alpha={\kappa^2_4 N^2\over 32\pi^2}\, , \quad
  \tilde{b_0}={2\over R^2}b_0\, .
\eeq
The equation (\ref{Freed-2}) is then solved
at first with respect to $\lambda$ as follows \footnote{ We notice here, 
 $$ \left({\dot{a}_0\over a_0}\right)^2+{k\over a_0^2}=\lambda\, . \label{bc-3-1} $$
}
\beq
   \lambda=\lambda_{\pm}\equiv {1\pm\sqrt{1-4\tilde{\alpha}^2
   \left(\Lambda_4/3+\tilde{\alpha}^2\tilde{b_0}^4\right)}\over 2\tilde{\alpha}^2}
\eeq
In the following,
we solve the above equations for $\lambda=\lambda_-$ and  $\lambda=\lambda_+$ separately. 
The solutions $a_0(t)$ of the
two equations can be connected at a singular point $a_0=a_0^{min}$, which is given below.

\section{Solution of $\lambda=\lambda_-$}

\subsection {For $\Lambda_4<0$}

The main purpose here is to study the state of the
theory in the region of $\lambda <0$, which is realized
for $\Lambda_4 <0$, since a phase transition is expected there. The 
case of $\Lambda_4 >0$ is studied afterward. 

The case of $\lambda<0$ is realized for the
solution, $\lambda=\lambda_-$ with negative $\Lambda_4 (=-|\Lambda_4|)$.
In this case, the equation to be solved is written as,
In this case, the equation to be solved is written as,
\beq\label{F-Eq-1}
  \left({\dot{a}_0\over a_0}\right)^2+{k\over a_0^2}= 
{1-\sqrt{1+4\tilde{\alpha}^2 |\Lambda_4|/3-4\tilde{\alpha}^4\tilde{b_0}^4 
   }\over 2\tilde{\alpha}^2}  \, .
\eeq
We notice $\tilde{b_0}^4={4C\over R^2 a_0^4}$.
And, here we suppose $k=-1$ to cover the region of negative $\lambda$.
For the solution $a_0(t)$ of this equation, the value of $\lambda$ varies in the range
\beq\label{co-3}
  {1-\sqrt{\tilde{\Lambda}_4}\over 2\tilde{\alpha}^2}\leq \lambda \leq{1\over 2\tilde{\alpha}^2}\, ,
  \quad \tilde{\Lambda}_4=1+4\tilde{\alpha}^2 |\Lambda_4|/3
\eeq
The negative $\lambda$ is found for
\beq\label{co-01}
   |\Lambda_4|/3>\tilde{\alpha}^2\tilde{b_0}^4\, .
\eeq
On the other hand, 
for the case,
\beq\label{co-11}
   |\Lambda_4|/3< \tilde{\alpha}^2\tilde{b_0}^4 \leq  {1\over 4 \tilde{\alpha}^2}+|\Lambda_4|/3\, ,
\eeq
we found positive $\lambda$. The right hand side of the inequality (\ref{co-11}) denotes the condition for the
reality of $\lambda_-$, and it provides 
the minimum of the allowed $a_0$.
It is given as 
\beq\label{a0-min}
  a_0^{min}=\tilde{\alpha} \left({16C\over R^2\tilde{\Lambda}_4}\right)^{1/4}\, .
\eeq
This point is called as "sudden singularity" 
\cite{Awad} since the scalar curvature diverges
at this point due to the fact that $\ddot{a}_0=\infty$ at this point as shown below.

We should notice that the left hand inequality of (\ref{co-11}) gives a constraint on $a_0$
from its upper side, but it is not the maximum value of possible $a_0$ determined by the Friedmann
equation. It is given by (\ref{F-Eq-1}) as shown below.

\vspace{.6cm}
\subsubsection{Solution of (\ref{F-Eq-1})}
\vspace{.3cm}

First, we consider the above equation (\ref{F-Eq-1}) by rewriting it into the following form,
\bea
  &&{1\over 2} {\dot{a}_0^2}+V_-(a_0)=0\, ,  \label{F-Eq-10}  \\ 
  V_-(a_0)&=&{k\over 2}-{1-\sqrt{1+4\tilde{\alpha}^2 |\Lambda_4|/3-4\tilde{\alpha}^4\tilde{b_0}^4 
   }\over 4\tilde{\alpha}^2}a_0^2  \, .  \label{F-Eq-11}
\eea

The value of the solution $a_0$ is restricted from below by $a_0^{min}$
as given by (\ref{a0-min}). The upper bound, denoted $a_0^{max}$,
is given as the turning point, where $\dot{a}_0=0$ or $V_-(a_0^{max})=0$.
Then from (\ref{F-Eq-10}) and (\ref{F-Eq-11}), we obtain it as 
\beq
  a_0^{max}=\left({1+\sqrt{1+4|\Lambda_4|\tilde{\alpha}^2(1+4C/R^2)/3}\over
        2|\Lambda_4|/3}\right)^{1/2}\, .
\eeq
Actually, we could see $V_-(a_0)>0$ for $a_0>a_0^{max}$, then there is no real
solution for $a_0>a_0^{max}$.

\vspace{.3cm}
Here, starting from $a_0=a_0^{min}$, we solve (\ref{F-Eq-1}) numerically
for appropriate values of $\Lambda_4$ and $C$
to see the characteristic properties of $a_0(t)$. 
A typical solutions is shown in the Fig. \ref{sol-m}.

We should notice here that, at $t=0$, $\dot{a}_0(t)$ is finite, but 
$\ddot{a}_0(t)=\infty$. This point is called as sudden singularity.
{So the scalar curvature is divergent as assured by the curve (c),
which represents $\ddot{a}_0(t)$,  in the Fig. \ref{sol-m}. 
Then
our equation would not be useful near this point, where higher order of curvature terms 
and /or the quantum gravity effect would be 
important there. This issue would be discussed in the future.

\begin{figure}[htbp]
\vspace{.3cm}
\begin{center}
\includegraphics[width=8cm]{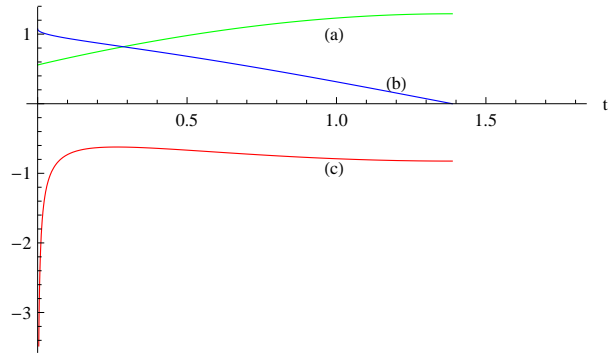}
\caption{Solution of $(a) a_0(t), (b) \dot{a}_0(t)$ and $(c) \ddot{a}_0(t)$ for $\lambda_{-}$, where $\Lambda_4/3=-1$, $C=0.03$ and $R=1$. }
\label{sol-m}
\end{center}
\end{figure}


Concentrating on the region
$a_0>a_0^{min}$, the following points are observed from the solution of (\ref{F-Eq-1}).

\begin{itemize}

\item It is well known that the solution $a_0(t)$ given by a periodic function
for constant, negative $\lambda$ and $C=0$. In the present case of $C\neq 0$, 
we find only one turning point at $a_0=a_0^{max}$ as seen from the Fig. \ref{sol-m}.
Namely $a_0$ starts from $a_0^{min}$ and arrives at this turning point, 
then it comes back to the starting point. 
After that, $a_0$ can not return to the $a_0^{max}$ since $\dot{a}_0<0$ there
when it comes back.
 
\item In the analysis, $a_0$ is restricted in the range
\beq
   a_0^{min} <a_0(t) \leq a_0^{max}\, .
\eeq
However, the solution doesn't oscillate between $a_0^{min}$ and $a_0^{max}$
since the point of $a_0^{min}$ is not a turning point but a singular point.
\end{itemize}

\vspace{.3cm}
\subsubsection{ Phase transition of SYM theory}

Using the solution of $a_0(t)$ given above, the time-dependence of the parameters
$b_0$ and $r_0$ is shown in the left of the Fig. \ref{r0-b0}. We find their cross point
where the phase transition occurs for the case of constant $a_0$. In the present case,
as explained below, 
the transition point is given by the zero point of the curve (c) in the same figure, and
it deviates slightly from the point of $b_0=r_0$.

\begin{figure}[htbp]
\vspace{.3cm}
\begin{center}
\includegraphics[width=7cm]{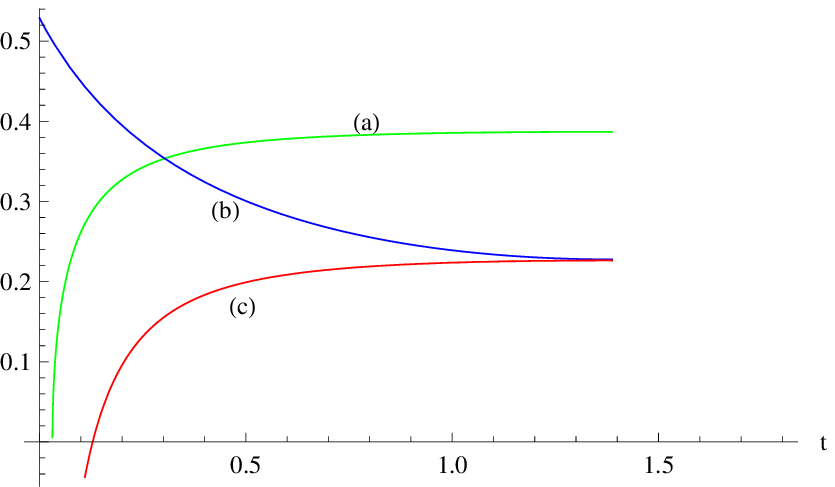}
\includegraphics[width=7cm]{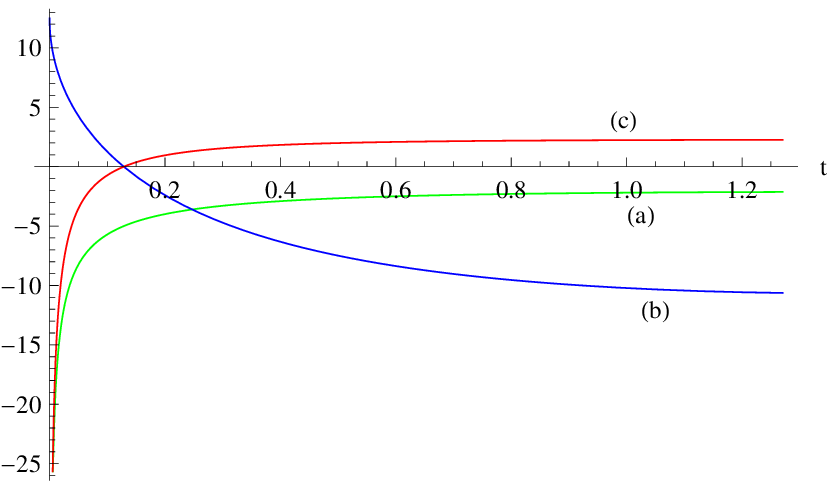}
\caption{Left;  Time dependence of $(a) r_0, (b)~ b_0 $ and (c) $n_an_b-n_c$. Right; Time dependence of $(a) 10/x_-, (b) 10^2/ x_+$ and (c) $10^2\times ( n_an_b-n_c)$ for $\lambda_{-}$, where $\Lambda_4/3=-1$, $C=0.03$ and $R=1$. 
 }
\label{r0-b0}
\end{center}
\end{figure}

The transition point is determined here as follows.
\begin{itemize}
\item (1) For each solution $a_0(t)$, we search for the horizon, which is given as a zero point of $\bar{n}(r,t)$, which is given by Eq. (\ref{sol-11}),
with respect to $r$. 
\item (2) When such a zero point could (not) be found, we can say that the phase of the state of the SYM theory
is in the deconfinement (confinement) phase. 
\item (3) In this way, we find a point, where the horizon disappears, as the transition point 
from deconfinement to the confinement phase.

\item (4) Performing the above procedure for different parameters with the same initial condition, we
find a critical curve in the $b_0-r_0$ plane. It is shown in the Fig. \ref{Phase-diagram}.
\end{itemize}
\vspace{.3cm}
In the step (1) of the above procedure, the zero of the numerator of $\bar{n}(r,t)$ in Eq. (\ref{sol-11}) is 
found by rewriting it as the quadratic equation of $x\equiv R^2/r^2 >0$, and the solutions are given as
\beq
   x_{\pm}={n_a+n_b\pm\sqrt{(n_a+n_b)^2-4(n_an_b-n_c)}\over 2(n_an_b-n_c)}\, ,
\eeq
where
\beq
  n_a={\lambda R^2\over4}\, , \quad n_b={\lambda+\dot{\lambda}{a_0 \over\dot{a}_0} \over 4} R^2\, , \quad n_c=\tilde{c}_0\, .
\eeq
Then we search for positive solution of $x_{\pm}$ as a horizon of the 5D metric (\ref{10dmetric-2-1}).

At the beginning, or at small $a_0$, a horizon can be observed at large $r$ (small $x_-$). 
In other words, the SYM system is in a very hot plasma phase. Then, we expect that the horizon
approaches to $r=0$ with growing $a_0$
and disappears at the transition point. 
We find that this transition point is realized at the point of
\beq\label{trans-p}
  n_an_b-n_c=0\, \quad {\rm or} \quad 
r_0=\left(1+{\dot{\lambda} a_0\over\lambda \dot{a}_0}  \right)^{-1/4} b_0\, .
\eeq
where the second equation is given in (\ref{critical-L1}) in the second section.
An example is shown by the right hand side figure of Fig. \ref{r0-b0}, where the transition point is found
at $t\sim 0.12$. Then the values of $b_0$ and $r_0$ are found from the solution shown in the left hand side
figure. We notice that this point shows a definite shift from the point of $b_0=r_0$.
After all, we find the critical curve (b) in the $b_0-r_0$ plane as shown in the Fig. \ref{Phase-diagram}.

\vspace{.3cm}
\noindent{\bf Tension of $q\bar{q}$ potential}
\vspace{.3cm}

We notice that the above transition point is also assured from the potential between the quark and antiquark.
In the confinement phase, a finite tension ($\tau_{q\bar{q}}$) for the linear potential
appears in the confinement phase.  
The tension $\tau_{q\bar{q}}$ is given by \cite{EGR}
\beq
    \tau_{q\bar{q}}={n_s(r^*)\over 2\pi\alpha'}\, , \quad n_s(r)=\left({r\over R}\right)^2\bar{n}\bar{A}\, ,
\eeq
where $r^*$ denotes the minimum point of $n_s(r)$. 
A typical calculation for $n_s$ is shown in the Fig. \ref{nstring}.
From this figure, we can see that the transition point coincides with the one given above
by observing the horizon.

\begin{figure}[htbp]
\vspace{.3cm}
\begin{center}
\includegraphics[width=12cm]{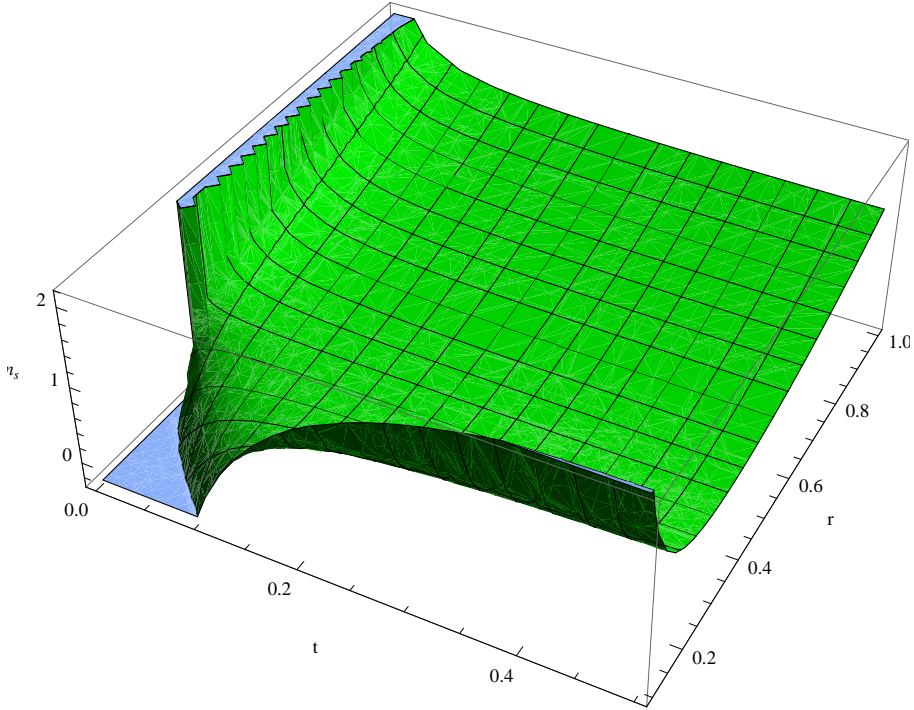}
\caption{Time dependence of $ n_s(r,t) $ for $\Lambda_4/3=-1$, $C=0.03$ and $R=1$. 
 }
\label{nstring}
\end{center}
\end{figure}

\vspace{.3cm}
\noindent{\bf Equation of state, $w$; Instead of $T$}
\vspace{.3cm}

The two critical lines (a) and (b) in the Fig. \ref{Phase-diagram} are definitely different.
This difference is reduced to the correction factor,
$\left(1+{\dot{\lambda} a_0\over\lambda \dot{a}_0}  \right)^{-1/4}$,
appeared in the case of time-dependent scale factor $a_0(t)$. 
On the other hand, we could find a quantity which characterizes the critical point by its special value
even if the time-dependent $a_0(t)$ is considered. So we can use this quantity like the 
temperature to fix the state of the system.

Such a nice quantity is the ratio $w(=p/\rho)$ of the pressure $p$ and energy density $\rho$.
This is called as equation of state (EOS).
Actually, at the critical point, we find
\beq\label{critical-p}
   w=-{1\over 3}\, .
\eeq
This is proved analytically by using (\ref{rho})-(\ref{density}) and (\ref{trans-p}). 
We should notice that the relation (\ref{critical-p}) is satisfied
at the critical point independently of various parameters. 
Furthermore, we can see the relation (\ref{critical-p}) for the case of constant $a_0$
at the critical point $r_0=b_0$. This point has not been discussed in the previous analysis.

In the present model, after the phase transition the temperature does not change
from $T=0$. However the state of the SYM system continues to change with time since the other quantities
like $\rho$ and $p$ then also $w$ change. 
This fact implies that the dynamical properties of the
system might be characterized by $w$ in the present case rather than by the temperature $T$.

\vspace{.3cm}
\noindent{\bf  A speculation on the critical point $w=-1/3$ from Virial theorem}
\vspace{.3cm}

When $a_0(t)$ varies with time, for small $a_0$ or for high density, we find
$w>-1/3$ and a horizon is seen. Afterward, 
$a_0$ increases and the horizon approaches to $r=0$ and disappears
for $w<-1/3$ in the large $a_0$ region. What has happened at the critical point, $w=-1/3$?

\vspace{.3cm}
For $C=0$, there is no radiation and only the negative $\Lambda_4$ exists. This case has been examined
in \cite{GIN2}, and we find the SYM theory is in the confinement phase with linear quark potential. In the case of $C>0$,
the above vacuum state is excited to a state with a radiated SYM fields, gluons. 

These excited gluons can be observed as the thermal radiation in the 
deconfinement phase where the screening effect for the confinement force is overwhelming. Here we replace the
strength of the screening effect by a force with the following form of potential,
\beq
   V_{screen}=aL^{\alpha}\, ,
\eeq
where $a$ and $\alpha$ are some constants and $L$ denotes the distance between the gluons (or 
between the quark and anti-quark). {This force would be repulsive.}

Next, we consider the virial theorem for the gluon (thermal-) system, and we obtain
\beq
    \langle K\rangle ={\alpha\over 2}  \langle V\rangle\, ,
\eeq
where $K$ and $V$ denote the kinetic- and potential- energy for the gluons respectively. 
The vacuum expectation values for them are taken quantum mechanically. The equation of state is related to them as
\beq
  w={\langle K\rangle- \langle V\rangle\over \langle K\rangle+ \langle V\rangle}
   ={\alpha-2 \over \alpha +2}\, .
\eeq
From this, we find $\alpha=1$ for $w=-1/3$. The system is in the confinement phase
for $\alpha <1$. This implies that, in the region of $w<-1/3$,
the confining potential ($V=\tau L$  \cite{GIN2}) overwhelms 
the screening (oe thermal) effect at large $L$. As a result, the system should be in the confinement
phase for $w<-1/3$.

It should be noticed that this is true when the system is in confinement phase for $C=0$.
When the system is in deconfinement phase for $C=0$, the system remains in a deconfinement
phase even if $w<-1/3$.

\vspace{.2cm}
\noindent{\bf Behaviours of $\rho$ and $p$}
\vspace{.2cm}

It is necessary to see the time-dependence of the physical quantities near the 
transition point. We show the energy density $\rho$, pressure $p$ and the parameter $w$,
in the Fig. \ref{eos} in the present case. 
Unexpectedly, we can not see
any sign of the phase transition in these quantities themselves since 
they all change very smoothly with time.

\begin{figure}[htbp]
\vspace{.3cm}
\begin{center}
\includegraphics[width=7cm]{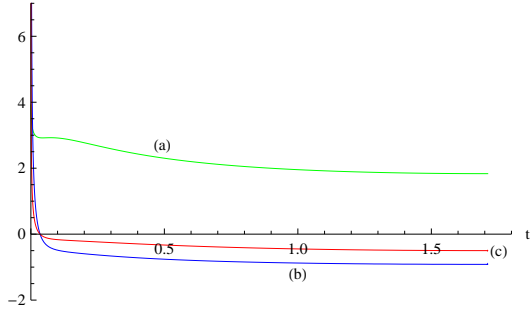}
\caption{ Time dependence of $(a) \rho, (b)  p $ and (c) $ w=p/\rho$ for $\lambda_{-}$, where $\Lambda_4/3=-1$, $C=0.03$ and $R=1$.  }
\label{eos}
\end{center}
\end{figure}

\vspace{.3cm}
\subsection{For the case of $\Lambda_4 >0$}

For $\Lambda_4 >0$, $\lambda_-$ is positive for all the region of $a_0>a_0^{min}$. 
In this case, the Friedmann equation to be solved is written 
for the case of $\lambda=\lambda_-$ as,
\beq\label{F-Eq-1-}
  \left({\dot{a}_0\over a_0}\right)^2+{k\over a_0^2}= 
{1-\sqrt{1-4\tilde{\alpha}^2 \Lambda_4/3-4\tilde{\alpha}^4\tilde{b_0}^4 
   }\over 2\tilde{\alpha}^2}  \, .
\eeq
Although this equation has a solution for $k=-1,~+1$ and $0$, it is
solved here for $k=-1$ to compare the case of negative $\Lambda_4$ in the above sub-section 4.1.

For the solution $a_0(t)$ of this Friedmann equation (\ref{F-Eq-1-}), the value of $\lambda$ varies in the range
\beq\label{co-3-}
  {1-\sqrt{\tilde{\Lambda}_{4_-}}\over 2\tilde{\alpha}^2}\leq \lambda \leq{1\over 2\tilde{\alpha}^2}\, ,
  \quad \tilde{\Lambda}_{4_-}=1-4\tilde{\alpha}^2 \Lambda_4/3\, .
\eeq
We assure that $\lambda$ is positive. As for the minimum of the $a_0$,
it is given by 
\beq\label{a0-min-}
  a_0^{min}=\tilde{\alpha} \left({16C\over R^2\tilde{\Lambda}_{4_-}}\right)^{1/4}\, .
\eeq

\vspace{.6cm}
\noindent{\bf Typical solution of (\ref{F-Eq-1-})}
\vspace{.3cm}

\begin{figure}[htbp]
\vspace{.3cm}
\begin{center}
\includegraphics[width=8cm]{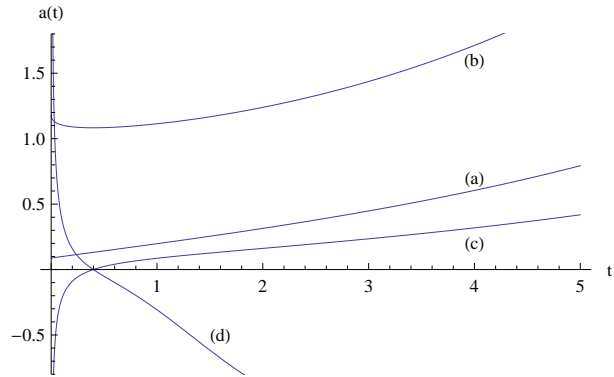}
\caption{Solution of $(a) a_0(t)/10, (b) \dot{a}_0(t)$, $(c) \ddot{a}_0(t)$, and $(d) (w-1/3)$ for $\lambda_{-}$, where $\Lambda_4/3=5\times 10^{-2}/(4\alpha^2)$, $C=0.03$ and $R=1$. }
\label{sol-p}
\end{center}
\end{figure}

The characteristic features of the solution of (\ref{F-Eq-1-}) can be read from the typical and numerical solution given
in the Fig. \ref{sol-p}. The equation has been solved by the intial condition, $a_0(0)=a_0^{min}$ as in the case of negative
$\Lambda_4$.

\begin{itemize}
\item The scale $a_0$ increases monotonically with almost constant velocity, $\dot{a}_0$. 
However, $\ddot{a}_0(t)$ changes its sign at $t=t_1$, so the expansion of the universe changes from decelerating to
the accelerating one at $t=t_1$. This implies that the dominant source of the expansion
changes from the radiation to the positive
dark energy at this time.
\item It would be meaningful to see the value of $w$ at this point. We find it as

\beq
     w(t_1)={1\over 3}\, .
\eeq 
This relation is also independent of the parameters of the theory as seen at the phase transition point above.
The meaning of this fact would be discussed in the future work.

\end{itemize}



\vspace{.3cm}
\noindent{\bf Emergence of the second horizon at $w=-1/3$}
\vspace{.3cm}

Next, we examine what happens at $w=-1/3$.
In the present case of positive $\Lambda_4$, $\lambda_-$ is positive at any point of $a_0(t)$.
As a result, the horizon represented by $x_-$ exists does not disappear at any time and as seen by the
curve (b) in the Fig. \ref{horizon-po}. 
In other words, the theory is
in the deconfinement phase at any time although the temperature decreases with increasing
$a_0$. In this sense, there is no phase transition to the confinement phase.

On the other hand,
we observe that $w$ decreases from a large positive value and crosses -1/3 at an appropriate
time, then decreases further and approaches to -1 for $t\to \infty$. It is interesting to see what happens in the 
present case after $w$ crosses $w=-1/3$. Our observation is summarized as follows.

\begin{figure}[htbp]
\vspace{.3cm}
\begin{center}
\includegraphics[width=8cm]{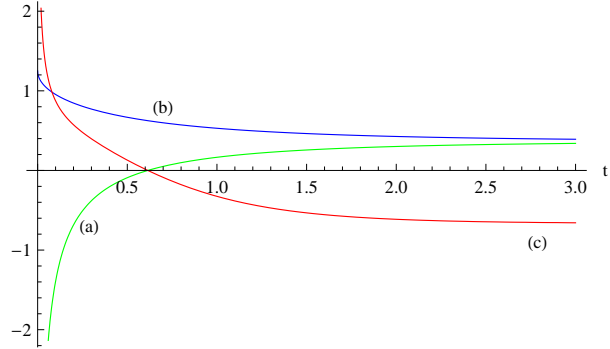}
\caption{Solution of $(a)~ 10/x_+, (b)~ 10/x_- $, $ (c)~ w+1/3 $ for $\lambda_{-}$, where $\Lambda_4/3=0.5/(4\alpha^2)$, $C=0.03$ and $R=1$. }
\label{horizon-po}
\end{center}
\end{figure}

\begin{itemize}
\item (i) There is no confinement deconfinement phase transition at 
this point as mentioned above. 
This is because of the 
fact that the theory for $C=0$, where the radiation disappears and $w=-1$, is in the deconfinement phase
as shown for the $\lambda >0$  case.
So the system should be continuing as the deconfinement phase even if the state is in the region
$-1/3>w(>-1)$. 

\item (ii) While the horizon ($x_-$) observed for $w>-1/3$ smoothly varies 
as a horizon with positive value. On the other hand,
when $w$ crosses $w=-1/3$, the value of $x_+$ changes its sign to positive. 
This implies that the second horizon, $x_+$
appears for $w>-1/3$. Then we could say that
in the holographic sense, two theories are explicitly
described by the dual 5D geometry for $-1<w<-1/3$.
{Two boundaries for each horizon} exist in this case, so we will find
two 4D field theories on those boundaries. A similar case has been discussed in \cite{Mar}.
However, the two CFT would be non-interacting
since they are not causal due to the two horizons between the boundaries \cite{MST}.
Similar situation has been found and discussed
in the case of negative $\lambda$ \cite{GNI13}.
On this point, we will discuss
in the future work.
\end{itemize}

\vspace{.3cm}
\noindent{\bf Instanton solution for $k=+1$} 
\vspace{.3cm}

Up to here, we have examined the case of Lorentzian time, and we find one turning point
for the case of $\Lambda_4 <0$. For $\Lambda_4 >0$, we can find a Euclidean time solution
with two turning points for $k=1$. Barvinsky, Kamenshchik, and Nesterev \cite{Barv}
have used this solution to construct a new hill-top inflation
scenario. In this sense, this type of solution is very interesting, but we will discuss
on this solution in the future. 


\section{Solution of $\lambda=\lambda_+$}

At the point of $a_0^{min}$, $\lambda_-$ takes its maximum value ${1\over 2\tilde{\alpha}^2}$. The value of 
$\lambda$ larger than this maximum of $\lambda_-$ is 
realized by $\lambda_+$. 
So the solution of the equation solved here covers the side of larger $\lambda$ than the case
of $\lambda=\lambda_-$.

In the present case, $\lambda$ is always positive, so we can take any value of $k$, $k=0,1,-1$.
In any case, $a_0$ grows exponentially with the $t$ at enough large $t$, 
where $\lambda$ is positive and almost constant as shown in the Fig. \ref{sol-pl}. The typical solution is 
shown in the left of
the Fig. \ref{sol-pl}. In this case, there is no turning point, so the Universe continues to expand 
exponentially forever.

\begin{figure}[htbp]
\vspace{.3cm}
\begin{center}
\includegraphics[width=7cm]{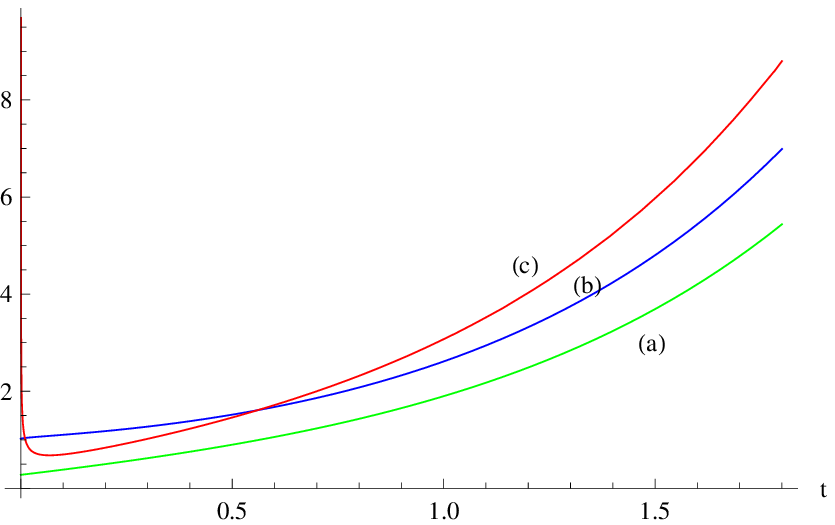}
\includegraphics[width=7cm]{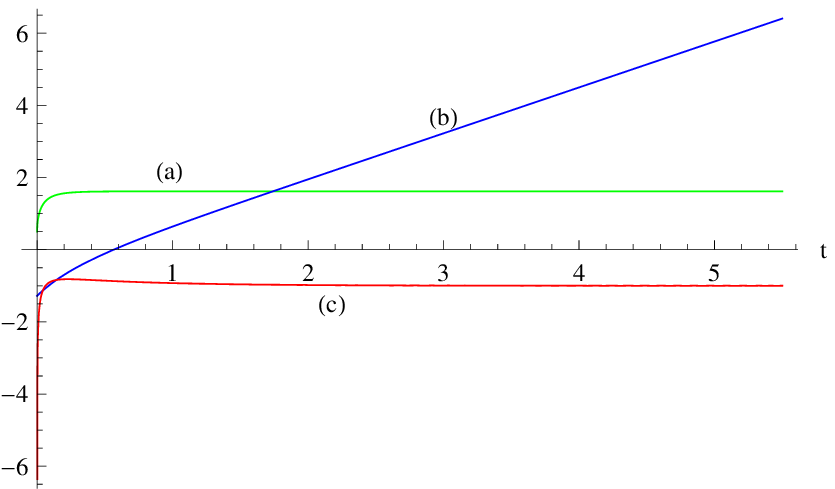}
\caption{Left; Time dependence of (a) $a_0(t)$, (b) $\dot{a}_0(t)$ and (c) $\ddot{a}_0(t)$ for $\lambda_{+}$, where $\Lambda_4/3=-1$, $C=0.03$ and $R=1$. 
Right; Time dependence of (a) $\lambda_{+}$, (b) $\ln a_0(t) $ and (c) $w=p/\rho$ }
\label{sol-pl}
\end{center}
\end{figure}

Some important observations with respect to $w$ are summarized as follows.

\begin{itemize}
\item First, we observe that $w$ increases from $-\infty$ to -1 with increasing
time. The state is restricted to the region of $w<-1$, where the state 
is called as the phantom state since the kinetic energy is negative. How do we interpret
this fact here is an open problem.

\item As mentioned above, $w(<-1)$ does not arrive at $-1/3$. 
As a result, the number of the horizon does not
changes. We observe two horizons at any t. See Fig. \ref{horizon-po-Lam}

\end{itemize}

\begin{figure}[htbp]
\vspace{.3cm}
\begin{center}
\includegraphics[width=8cm]{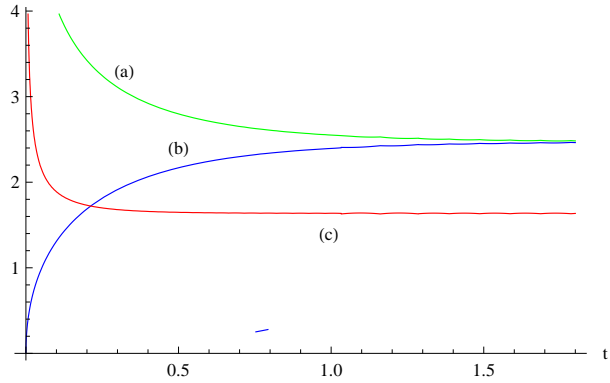}
\caption{Solution of $(a)~ x_+, (b)~ x_- $, $ (c)~ 10(n_an_b-n_c) $ for $\lambda_{+}$ and $k=-1$, where $\Lambda_4/3=-1/(4\alpha^2)$, $C=0.03$ and $R=1$. }
\label{horizon-po-Lam}
\end{center}
\end{figure}

\vspace{.3cm}
\noindent{\bf About the Connectedness of $\lambda_-$ and $\lambda_+$ at
$a_=a_0^{min}$}
\vspace{.3cm}

As pointed out above, the solutions of $\lambda=\lambda_-$ and $\lambda=\lambda_+$ are
connected at the sudden singularity, namely at the point of $a_0=a_0^{min}$. 
The scalar curvature is divergent at this point. So it is not
simple to connect the two solutions due to the singularity. As for this problem, there is an interesting approach
proposed by Awad \cite{Awad}. This idea is very interesting to make an inflation scenario, however,
we would postpone to discuss this issue to the future work.

\section{Summary and Discussion}

In AdS/CFT, boundary gravity decouples from the bulk, and the boundary metric becomes nondynamical. 
Hence, the time-dependent scale factor $a_0(t)$ in a FRW space-time on the boundary can not be determined
self-consistently from the equations of motion of the bulk gravity theory, which in our case is a truncation of 10D supergravity.  
On the other hand, the bulk solution induces energy and momentum sourcing gravity on the boundary via the usual AdS/CFT dictionary for the metric. For example, a finite temperature state in the bulk (an AdS-Schwarzschild black hole), will induce a radiation energy density and a corresponding pressure on the boundary. Hence, in order to obtain a self-consistent holographic cosmological evolution, we need to impose the dynamics of the scale factor $a_0(t)$ by hand. 

In this work we proposed a way to do so by imposing the Friedmann equations for the boundary metric which relate energy density and pressures of the dual field theory to the time-development of the CFT by giving a solution for the scale factor $a_0(t)$. 
We hence solved the 4D boundary Einstein equations coupled to the SYM 
theory energy-momentum tensor $\langle T_{\mu\nu}\rangle$ and to a boundary cosmological constant $\Lambda_4$. The energy momentum tensor $\langle T_{\mu\nu}\rangle$ itself is obtained from the holographic renormalization procedure in the the 5D reduction of type IIB 10D supergravity. 
By using the $a_0(t)$ solved in this way we were able to find a time dependent 5D gravitational background, which consistently describes the time dependence 
of the state of $\cal N=4$ SYM theory in the FRW space-time via the AdS/CFT correspondence..  
The solution is characterized by two free parameters, $\Lambda_4$ and $C$,
the 4D cosmological constant and the dark radiation constant of the SYM fields. 
These two parameters control the dynamical properties of the SYM theory. 

We then proceeded to analyze the phase structure of the SYM theory for different cases of these parameters. Our general finding is that negative $\Lambda_4<0$ drives the theory to a Wilson loop confinement phase. 
On the other hand, the dark radiation $C$ counteracts this tendency by the screening of the confinement force. 
Hence these competing effects lead to a confinement-deconfinement transition. 
The same phenomenon had been observed for the case of slowly varying $a_0$ 
already in \cite{EGR}-\cite{GN13}. 

\vspace{.3cm}
We find that the solution for  
$a_0(t)$ has two branches, $\lambda=\lambda_-$ and $\lambda=\lambda_+$. These two branches arise from solving a quadratic equation for $\lambda$. \footnote{
We should notice here that $\lambda_{\pm}$ is not $\Lambda_4$ but it is given by 
$\Lambda_4$ and the dark radiation as shown above.
} 
Both solutions have a minimum $a_0^{min}$ of the scale factor $a_0(t)$, and both  
branches $\lambda_{\pm}$ meet at this point. 
This point turns out to be singular 
since the acceleration $\ddot{a}=\infty$ diverges. At this point, the classical Friedmann equation breaks down, and quantum gravitational effects have to be taken into account to resolve the singularity, which is beyond the scope of this work. 
   
The first branch $\lambda=\lambda_-$ itself separates into two cases depending on the sign of $\Lambda_4$. 
For negative $\Lambda_4$, the region of negative values of $\lambda_-$ is covered. The Friedmann equation 
is then solved for hyperbolic ($k=-1$) three-dimensional spatial topology and we find that the solution $a_0(t)$ increases with time and arrives at a maximum turning point with $\dot{a}_0=0$ and $a_0^{max}$. After that, $a_0$ turns back to the singular point $a_0^{min}$. An important phenomenon in this case
is that, before arriving at $a_0^{max}$, we find a phase transition of the SYM theory from
the (Wilson loop) deconfinement to the confinement phase at a critical time $t=t_c$. This transition happens exactly when the horizon zero in $g_{tt}$,
which is present in the deconfinement 
phase at small $a_0(t<t_c)$, disappears. We have observed that at this transition point, the value of
the ratio $w=p/\rho$ of the pressure to the energy density 
 is exactly $-1/3$. This value is in particular independent of the two free parameters of the theory $C$ and $\Lambda_4$. The meaning of this value 
$w=-1/3$ has been discussed from the viewpoint of the Virial equilibrium.  
We would like to stress that the energy-momentum tensor components $\rho$ and $p$ (and hence also $w$) 
vary smoothly across the transition, which hence only shows in the Wilson loop potential. 

In the case of $\Lambda_4 >0$, $\lambda$ is always positive, and we can solve the Friedmann equation for 
any value of the spatial curvature $k$, namely for $k=-1,~+1$ 
and $0$. In order to compare with the case of $\Lambda_4 <0$, we have first  
examined the case of negative curvature, $k=-1$. In this case, the solution $a_0$
increases monotonically and has no turning point. It furthermore has the following properties: 
First, the acceleration $\ddot{a}(t)$ changes its sign 
from positive to negative at an appropriate time. While we find that  
$w=+1/3$ is realized at this changing point 
independently of the parameters, its physical interpretation is an open question here. On the other hand, we find no phase transition at the point of $w=-1/3$ since the horizon zero in $g_{tt}$ remains present for all regions of $a_0(t)$. Thirdly, we noticed that a second horizon appears for $w<-1/3$, an interesting observation whose meaning we plan to investigate in a future work. Finally, for $k=+1$ we found a Euclidean solution with two turning points. This solution has been used in \cite{Barv} 
as an instanton solution to drive an inflation scenario. 
In our case such solutions exist for special values of the parameters, $\Lambda_4$ and $C$. 
We postpone to discuss about this kind of solution related to the inflation scenario and the quantum cosmology.

\vspace{.3cm}
For the second branch, $\lambda=\lambda_+$, the solution $a_0$ increases monotonically
with time in all cases and never has a turning point. At enough large $a_0$, it expands exponentially as expected since $\lambda_+$ is positive and grows $a_0(t)$. 
The value of $w$ increases monotonically from $-\infty$ to -1, so
$w<-1$ at any $a_0$. So the matter is in a phantom phase.
We find always two horizons in this case. 

\vspace{.3cm}
Of course, the time dependence of the solutions $a_0$ obtained here depends on the 4D gravity model on the boundary as well as on the 5D bulk theory and the chosen solution therein. In the present case, it is constructed by the Einstein-Hilbert action with a cosmological constant
$\Lambda_4$ and SYM theory. In general, other matter fields might 
be included, or the 4D Einstein equations could be modified. 
For example, higher order curvature terms may be needed near the "sudden" singularity 
to resolve it. In this sense, some parts of the above results will be model dependent, but the chosen model, a otherwise conformal ground state in the bulk with only temperature turned on as well as the prudent choice of the standard two-derivative Einstein-Hilbert action coupled minimally to energy and momentum as well as a cosmological constant on the boundary is the simplest reasonable choice. Also, the results obtained here would be remained as useful clue when we perform the analysis of more complicated models.

\vspace{.3cm}
\section*{Acknowledgments}

The authors would like to thank M. Ishihara for useful discussions and comments. K. G thanks 
J. Erdmenger for encouragement.



\newpage
\end{document}